\documentclass{article}

\usepackage{graphicx,psfrag,epsf}
\usepackage{arxiv}
\usepackage[utf8]{inputenc} 
\usepackage[T1]{fontenc}    
\usepackage{hyperref}       
\usepackage{url}            
\usepackage{booktabs}       
\usepackage{amsfonts}       
\usepackage{nicefrac}       
\usepackage{microtype}      
\usepackage{lipsum}
\usepackage{natbib}
\usepackage{amsmath}
\usepackage{amssymb}
\usepackage{amsthm}
\def\T{{ \mathrm{\scriptscriptstyle T} }}
\def\0{{\bf 0}}
\def\uy{{\bf y}}
\def\ud{{\bf d}}
\def\ue{{\bf e}}
\def\uz{{\bf z}}

\def\uZ{{\bf Z}}
\def\uP{{\bf P}}
\def\uR{{\bf R}}
\def\uM{{\bf M}}
\def\uS{{\bf S}}
\def\uI{{\bf I}}
\def\ualpha{{\boldsymbol \alpha}}
\def\ugamma{{\boldsymbol \gamma}}
\def\ueta{{\boldsymbol \eta}}
\def\utheta{{\boldsymbol \theta}}
\def\uSigma{{\boldsymbol \Sigma}}
\newtheorem{theorem}{Theorem}

\title{Bayesian causal inference with some invalid instrumental variables}
\author{
  Gyuhyeong Goh\\
  Department of Statistics\\
  Kansas State University\\
  Manhattan, KS 66506\\
  \texttt{ggoh@ksu.edu} \\
   \And
  Jisang Yu\\
  Department of Agricultural Economics\\
  Kansas State University\\
  Manhattan, KS 66506\\
  \texttt{jisangyu@ksu.edu} \\
}

\begin{document}
\maketitle

\begin{abstract}
In observational studies, instrumental variables estimation is greatly utilized to identify causal effects. One of the key conditions for the instrumental variables estimator to be consistent is the exclusion restriction, which indicates that instruments affect the outcome of interest only via the exposure variable of interest. We propose a likelihood-free Bayesian approach to make consistent inferences about the causal effect when there are some invalid instruments in a way that they violate the exclusion restriction condition. Asymptotic properties of the proposed Bayes estimator, including consistency and normality, are established. A simulation study demonstrates that the proposed Bayesian method produces consistent point estimators and valid credible intervals with correct coverage rates for Gaussian and non-Gaussian data with some invalid instruments. We also demonstrate the proposed method through the real data application.
\end{abstract}


\section{Introduction}\label{sec:1}
In observational studies, identifying the causal effect of an exposure variable on the outcome of interest has been a great interest and a challenge for researchers in various disciplines. Utilizing instrumental variables (IV) is one of the popular methods to obtain the properly identified causal estimates and the robustness of the results. For example, in epidemiology and genetics studies, increasing usage on Mendelian randomization demand in-depth understanding of the IV estimations \citep{davey2003mendelian,smith2004mendelian,didelez2007mendelian}. Similar developments have occurred for observational studies in economics since the Leamer critique \citep{leamer1983let,angrist1996identification,angrist2010credibility}, which diagnoses that many empirical studies suffer from a lack of robustness. 

Better data availability for observational studies along with the credibility revolution \citep{angrist2010credibility} has stimulated the methodological innovations in causal inference studies including the innovations on the IV estimation, particularly in the context of the utilization of many instruments. Recent literature has developed several IV estimation techniques that focus on sparsity estimation problems to properly utilize many instrumental variables \citep{belloni2012sparse,belloni2013least,chernozhukov2015post,kang2016instrumental,windmeijer2018use}. While the majority of the studies in this area focuses on the use of variable selection methods to tackle weak instruments problems, few exceptions such as \citet{kang2016instrumental} and \citet{windmeijer2018use} develop penalized IV estimation methods that detect invalid instruments, which violates the exclusion restriction condition, and estimate the causal effect simultaneously. Our contribution lies on this recent development on estimating causal effects with many instruments when there are some possible invalid instruments.

In this paper, we propose a new Bayesian method that provides valid statistical inference for estimating casual effects when some instrumental variables violate the exclusion restriction, i.e. some instruments have direct impacts on the outcome of interest. Our proposed method has the following contributions. First, the proposed Bayesian estimator successfully accounts for model uncertainty in the selection of invalid instruments. The latest innovation in estimating causal effects in the presence of invalid instruments is to utilize penalized regressions so that the selection of invalid instruments and the estimation of the causal effect of interest can be done simultaneously  \citep{kang2016instrumental,windmeijer2018use}. The variances of the estimates on the causal effect of interest from these approaches can be underestimated because of the uncertainty in tuning parameter selections \citep{madigan1994model}. We contribute to the literature by addressing the model uncertainty issue by employing Bayesian model averaging for our Bayesian inference so that the model uncertainty (and the tuning parameters) associated with invalid instruments can be automatically integrated out. See \citet{hoeting1999bayesian} for a comprehensive review of Bayesian model averaging and \citet{lopes2014bayesian} for a review of Bayesian instrumental variables approaches.

The second contribution lies on the distributional assumption for data. The advantage of our proposed method is that it does not require any distributional assumption. Leveraging the equivalence between the two-stage least squares method and the generalized method of moments \citep{hansen1982large, andrews1999consistent}, we develop a Bayesian generalized method of moments approach that uses a pseudo-likelihood function constructed by moment conditions instead of a likelihood function that requires some parametric distributional assumptions. In addition, we show that the proposed method possesses model selection consistency from a Bayesian perspective. As a result, for sufficiently large sample sizes, our Bayesian estimator with a noninformative prior tends to be as good as the oracle two-stage least squares estimator obtained by knowing the true invalid instruments \textit{ex ante}.

Finally, our Bayesian generalized method of moments approach departs from the literature by handling the cases when the number of parameters is greater than the number of moment conditions. The idea of the Bayesian generalized method of moments was originally proposed by \citet{chernozhukov2003mcmc} and further developed by \citet{yin2009bayesian}. Theoretical properties of the Bayesian generalized method of moments including model selection consistency have been extensively studied \citep[e.g.][]{li2016oracle} and it has been applied to the context of IV estimation \citep[e.g.][]{liao2011posterior,kato2013quasi}. However, the IV estimations in the presence of  invalid instruments among many potential instruments, where the number of parameters is larger than the number of moment conditions, have not been explored by the existing Bayesian literature. This situation violates the regularity assumption of the existing studies on the Bayesian generalized method of moments. We mitigate this issue by imposing a constraint on the number of valid instruments, which is based on the identifiability condition of \citet{kang2016instrumental}, via the model prior distribution. The details are described in Section \ref{sec:3}. 

The IV problem can be illustrated as the following. Let $Y$ be a scalar outcome variable, $D$ be a scalar exposure variable, and $Z=(Z_1,\ldots,Z_p)^\T $ be a $p\times 1$ vector of instrumental variables. Denote the $n\times 1$ vector of the observed outcomes, the $n\times 1$ vector of the observed exposures, and the $n\times p$ matrix of the observed instruments by  $\uy=(y_1,\ldots,y_n)^\T$, $\ud=(d_1,\ldots,d_n)^\T$, and $\uZ=(\uz_1,\ldots,\uz_n)^\T$, respectively, where $(y_i,d_i,\uz_i)$, $i=1,\ldots,n$, are assumed to be $n$ independent realizations of a joint distribution of $(Y,D,Z)$. For each subject $i \in \{1,\ldots, n\}$, consider a simple outcome model, $y_i=\beta d_i+\varepsilon_i,$ where $\beta\in \mathbb{R}$ represents the causal effect of $d_i$ on $y_i$ and $\varepsilon_i$ denotes an independent realization of a random error $\varepsilon$. We assume that $y_i$, $d_i$, and $\uz_i$ $(i=1,\ldots,n)$ are centered around their sample means so that the intercept term is collapsed to zero in the outcome model. Here, we are interested in estimating the causal effect $\beta$.

In non-randomized observational studies, it is likely that the exposure $D$ is correlated with the error term $\varepsilon$ which causes an identification threat \citep{rubin1974estimating, rubin1978bayesian}. If $\text{cov}(D,\varepsilon)\neq 0$, the ordinary least squares estimator, $\hat{\beta}_{\text{ols}}=(\ud^\T \ud)^{-1} \ud^\T \uy$, is inconsistent since $\hat{\beta}_{\text{ols}}\to \beta+\text{cov}(D,\varepsilon)/\text{var}(D)$ in probability as $n\to \infty$. A common approach in such cases is to use two-stage least squares estimation by utilizing the instrument variables. For each $j\in \{1,\ldots,p\}$, $Z_j$, the $j$th component of the $p\times 1$ instrument vector $Z$, is called a \emph{valid} instrument if (i) there is no direct effect of $Z_j$ on $Y$, (ii) $Z_j$ is uncorrelated with $\varepsilon$, and (iii) $Z_j$ is associated with the exposure $D$. 

Suppose that a researcher treats all $Z_1,\ldots,Z_p$ as valid, then the na\"ive two-stage least squares estimator is defined as $\hat{\beta}_{\text{na\"ive}}=(\hat{\ud}_\uZ^\T \hat{\ud}_\uZ )^{-1} \hat{\ud}_{\uZ}^\T  \uy,$ where $\hat{\ud}_{\uZ}=\uP_\uZ \ud$ and $\uP_\uZ=\uZ (\uZ^\T \uZ)^{-1} \uZ^\T$. When the validity assumption for all instruments holds, $\hat{\beta}_{\text{na\"ive}}$ is consistent and equals to the oracle estimator, i.e., $\hat{\beta}_{\text{na\"ive}}\to \beta$ in probability as $n\to \infty$. The challenge arises, however, from the fact that the validity assumption for some instruments is often violated and the researcher may not know the set of invalid instruments. In this paper, our main objective is to develop a consistent Bayesian solution to IV estimation problems when invalid instruments are present.

The rest of the paper is organized as follows. In Section \ref{sec:2}, we illustrate IV estimation methods in the presence of invalid instruments and discuss challenges and limitations of the existing methods. In Section \ref{sec:3}, we propose a new Bayesian approach to IV estimation and establish some asymptotic properties of the proposed method. In Section \ref{sec:4}, we report the results from a simulation study to assess the finite-sample performance of the proposed method. In Section \ref{sec:5}, the proposed method is applied to a real data example from the seminal work of \citet{angrist1991does}. Section \ref{sec:6} concludes with some remarks. The appendix contains the proofs of our main theorems.

\section{Instrumental variables estimation with some invalid instruments}\label{sec:2}
Consider a generalization of the outcome model as follows:
\begin{eqnarray}\label{m:1}
y_i=\beta d_i+\uz_i^\T\ualpha +\epsilon_i,\quad i=1,\ldots,n,
\end{eqnarray}
where $E(\epsilon_i|\uz_i)=0$, $E(\epsilon_i^2|\uz_i)=\sigma^2_\epsilon$, $\epsilon_i$ is assumed to be independent of $\uz_i$, and $\ualpha=(\alpha_1,\ldots,\alpha_p)^\T$ represents the direct effect of $\uz_i=(z_{i1},\ldots,z_{ip})^\T $ on $y_i$ or the indirect effect generated by a correlation between $\uz_i$ and the unobservable confounding variables or the latent variables \citep{small2007sensitivity,kang2016instrumental,windmeijer2018use}. In the outcome model \eqref{m:1}, the validity of instrumental variables hinges on the exclusion restriction condition that $\ualpha=(0,\ldots,0)^\T$. In practice, particularly with many potential instruments, research designs may face the violation of the validity of instruments, i.e., some elements of $\alpha$ are possibly non-zero in the outcome model. 

\begin{figure}
\centering
\makebox{
\includegraphics[width=.45\textwidth]{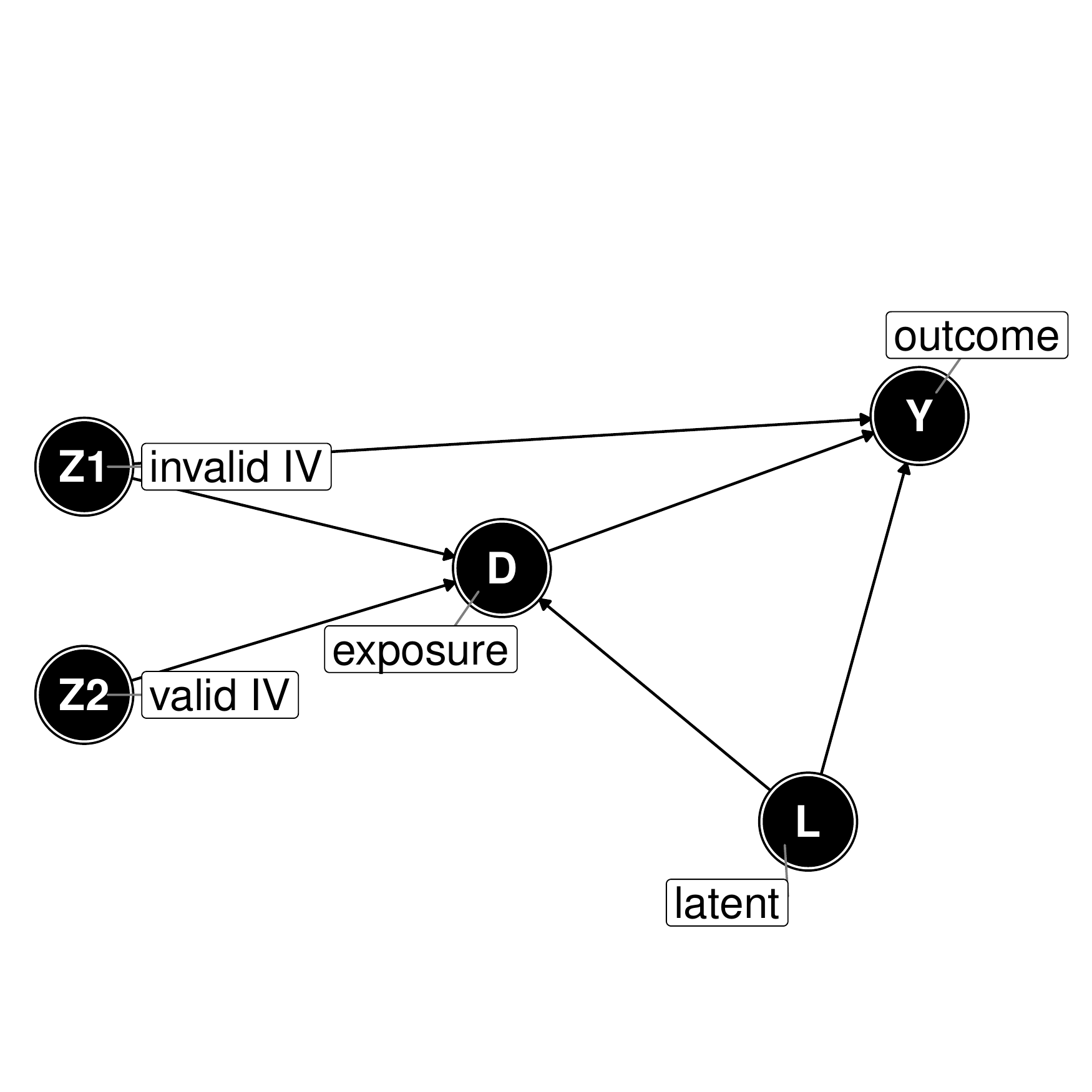}
\includegraphics[width=.45\textwidth]{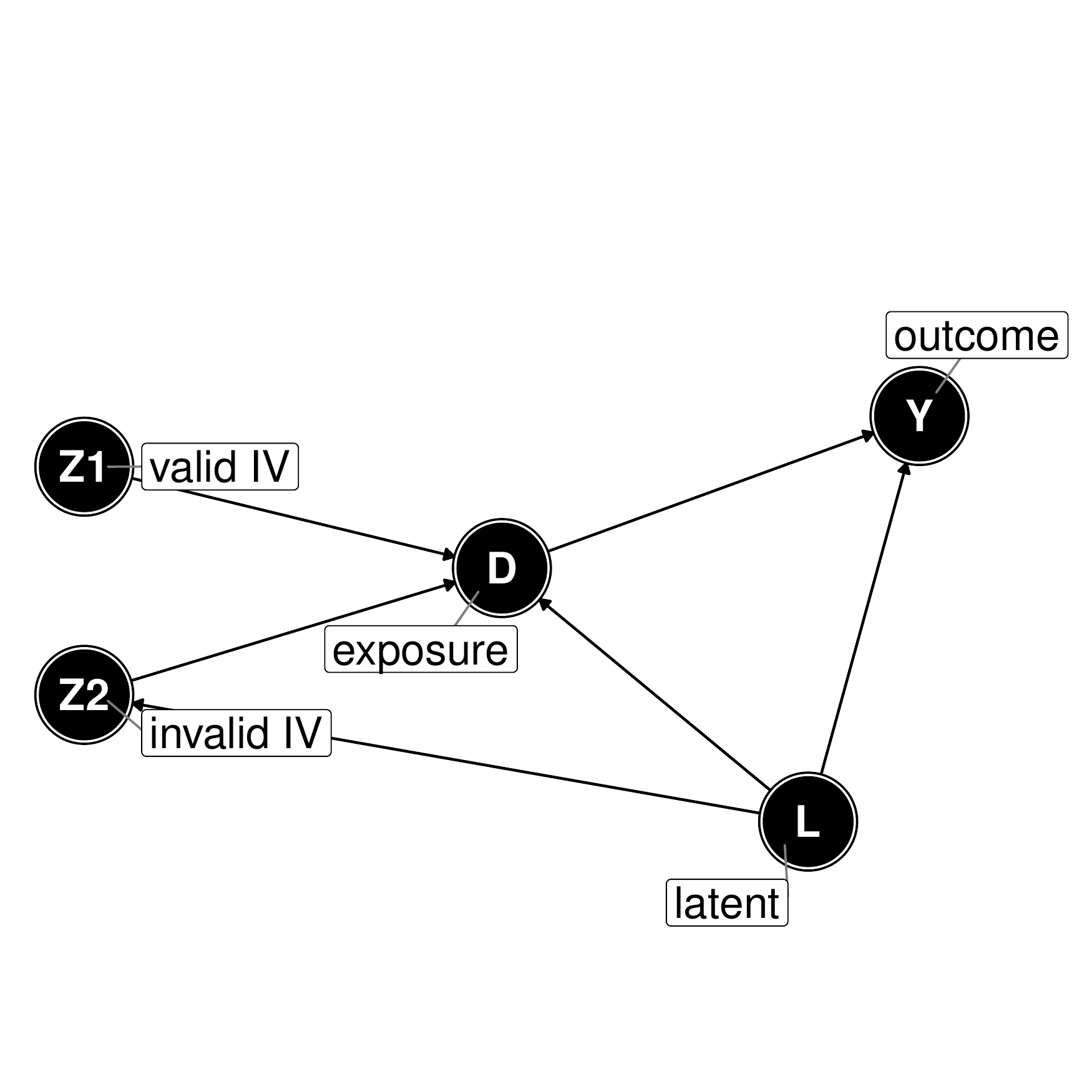}
}
\caption{Two examples of violation of the exclusion restriction. \label{fig:1}}
\end{figure}

Figure \ref{fig:1} illustrates two possibilities of violating the exclusion restriction. In the first example (left), \texttt{Z1} is an invalid IV due to the fact that \texttt{Z1} has a direct impact on the outcome. In the second example (right), \texttt{Z2} violates the condition that instruments are uncorrelated with the latent variables.

To formally define such circumstances, let $\omega=\{j:\alpha_j \neq 0\}$ be the index set of invalid instruments. If $\omega$ is a non-empty set, it can be shown that the na\"ive two-stage least squares estimator, $\hat{\beta}_{\text{na\"ive}}$, is inconsistent. Let $\uZ_{\omega}$ be a sub-matrix of $\uZ$ that is determined by the index set $\omega$. When $\omega$ is known and $\omega\neq \{1,\ldots p\}$, a consistent estimator for $\beta$, called the oracle two-stage least squares, can be obtained as $\hat{\beta}_{\text{oracle}}=(\hat{\ud}_{\uZ}^\T \uM_{{\uZ}_{\omega}} \hat{\ud}_{\uZ} )^{-1} \hat{\ud}_{\uZ}^\T  \uM_{\uZ_{\omega}}\uy$, where $\uM_{\uZ_{\omega}}= \uI_n- \uZ_{\omega}(\uZ_{\omega}^\T \uZ_{\omega})^{-1}\uZ_{\omega}^\T $. Under certain regularity conditions, the oracle two-stage least squares satisfies
\begin{eqnarray}\label{oracle:normality}
 \sqrt{n}(\hat{\beta}_{\text{oracle}}-\beta) \to \mathcal{N}(0,\sigma^2_{\text{oracle}}),
 \end{eqnarray}
in distribution as $n\to \infty$, where 
\begin{eqnarray*}\label{oracle:variance}
\sigma^2_{\text{oracle}}=\sigma^2_\epsilon\{E(Z^\T D) E(ZZ
^\T)^{-1} E(Z D)-E(Z_{\omega }^\T D) E(Z_{\omega }Z^\T_\omega)^{-1} E(Z_{\omega } D)\}^{-1},
\end{eqnarray*}
and $Z_{\omega}$ is a sub-vector of $Z$ corresponding to the indices in $\omega$.

However, implementing the oracle two-stage least squares estimation is mostly infeasible since $\omega$ is unknown in practice. Recent studies such as \citet{kang2016instrumental} and \citet{windmeijer2018use} introduce penalized least squares approaches to identify invalid instruments and estimate the causal effect simultaneously. A penalized least squares estimator is obtained by minimizing
 \begin{eqnarray}\label{pls}
 \|\uP_\uZ(\uy- \ud\beta-\uZ\ualpha)\|_2^2+\mathcal{P}_\lambda(\ualpha),
 \end{eqnarray}
over $(\beta,\ualpha)$, where $\|\cdot\|_2$ denotes the $L_2$-norm and $\mathcal{P}_\lambda(\cdot)$ is a penalty function with a regularization parameter, or tuning parameter, $\lambda> 0$. For example, \citet{kang2016instrumental} propose to use the $L_1$-penalty, also known as the lasso penalty, $\mathcal{P}_\lambda(\ualpha)=\lambda \| \ualpha\|_1$, where $\|\cdot\|_1$ denotes the $L_1$-norm. As the lasso penalty produces sparse solutions for $\ualpha$, variable selection and estimation can be done simultaneously in \eqref{pls}. 

Recently, however, \citet{windmeijer2018use} show that applying the lasso penalty to \eqref{pls} leads to an inconsistent estimator for $\beta$. As an alternative, they propose an adaptive lasso penalty constructed from a $\sqrt{n}$-consistent estimator of $\ualpha$. To formulate their idea, let $\hat{\gamma}_j$ and $\hat{\eta}_j$ be the $j$th elements of $\hat{\ugamma}=(\uZ^\T \uZ)^{-1} \uZ^\T \uy$ and $\hat{\ueta}=(\uZ^\T \uZ)^{-1}\uZ^\T \ud$, respectively. The adaptive lasso penalty is then defined as
$$\mathcal{P}_\lambda(\ualpha)=\lambda \sum_{j=1}^p |\alpha_j|/|\hat{\alpha}_{\text{m}j}|^\nu,$$
for a pre-specified $\nu>0$, where $\hat{\alpha}_{\text{m}j}$ is the $j$th element of $\hat{\ualpha}_{\text{m}}=(\uZ^\T \uZ)^{-1} \uZ^\T (\uy-\ud\hat{\beta}_{\text{m}}) $ and $\hat{\beta}_{\text{m}}={\rm median}\{\hat{\gamma}_1/\hat{\eta}_1,\ldots\hat{\gamma}_p/\hat{\eta}_p\}$, which is often called the median estimator \citep{han2008detecting}. 
\citet{windmeijer2018use} verify that the resulting adaptive lasso estimator of $\ualpha$ possesses variable selection consistency and asymptotic normality and, furthermore, that the limiting distribution of the adaptive lasso estimator for $\beta$ is equivalent to that of the oracle estimator in \eqref{oracle:normality}.

Although the penalized instrumental variables estimation enjoys large-sample properties and computational efficiency, there remain major limitations. First, the resulting estimates are sensitive to the choice of tuning parameter values. Hence, inappropriate selections of the tuning parameter often yield nonsensical estimates of the parameters of interest. Second, estimating variance of the penalized estimator is quite difficult in the IV estimation context. To alleviate this problem, the post-penalization estimator, that applies two-stage least squares to the model selected by the penalized least squares estimator, is commonly used in practice \citep{belloni2012sparse, chernozhukov2015post}. However, the success of post-penalization estimation strategy highly relies on the model selection performance of the penalized estimator. Third, the resulting sparse estimator ignores the uncertainty associated with model selection. \citet{madigan1994model} remark that ignoring such model uncertainty results in underestimation of the uncertainty about the parameter of interest. To address all the aforementioned issues, we propose a new Bayesian approach that provides a general framework for simultaneously selecting valid instruments and accounting for the uncertainties associated with variable selection and estimation. 

\section{Bayesian estimation}\label{sec:3}
As shown by \citet{hansen1982large, andrews1999consistent}, the two-stage least squares IV estimator is a special case of a generalized method of moments estimator. Using this relationship, we develop a new generalized method of moments framework that provides a Bayesian solution to identify the causal effects when there exist invalid instruments among the set of the instruments. The use of a Bayesian generalized method of moments approach enables us to avoid imposing any parametric distribution assumptions on the outcome model \eqref{m:1}. See \citet{yin2009bayesian, li2016oracle} for a general overview of the Bayesian generalized method of moments and also see \citet{liao2011posterior,kato2013quasi} for applications to nonparametric IV problems.

For a given $\omega$, let $\uR_{\omega}=(\ud,\uZ_{\omega})$ and $\utheta_{\omega}=(\beta,\ualpha_{\omega}^\T)^\T$, where $\ualpha_{\omega}$ is the sub-vector of $\ualpha$ corresponding to ${\omega}$. Using the moment condition of instrumental variables, $E(Z\epsilon)= \0$, we define the sample moment by $m_n(\utheta_{\omega})=n^{-1} \uZ^\T (\uy-\uR_{\omega}\utheta_{\omega})$ for given $\omega$. If $\omega$ includes all the true invalid instruments, then, under the regularity conditions of \citet{hansen1982large}, the asymptotic normality of the sample moment holds, i.e., 
\begin{eqnarray}\label{lim:dis}
\sqrt{n} m_n(\utheta_{\omega})\mid \utheta_{\omega},  \omega \to \mathcal{N}(0,\uSigma_m),
\end{eqnarray}
in distribution as $n\to \infty$, where $\uSigma_m=\sigma^2_\epsilon E(ZZ^\T)$. For given $\omega$, a consistent estimator of $\uSigma_m$ can be obtained as $\hat{\uSigma}_m=(n^{-1} \|\uy-\hat{\uy}_\omega \|_2^2)(n^{-1} \uZ^\T \uZ)$, where $\hat{\uy}_\omega=\uR_{\omega}(\uR_{\omega}^\T \uP_{\uZ} \uR_{\omega} )^{-1} \uR_{\omega}^\T  \uP_{\uZ} \uy$. 

Applying Slutsky's theorem with \eqref{lim:dis}, we define a pseudo-likelihood function as 
\begin{eqnarray}\label{like:1}
\tilde{f}( \uy \mid \utheta_{\omega}, \omega)=\frac{1}{(2\pi \hat{\sigma}^2_{\epsilon})^{p/2} |n^{-1} \uZ^\T \uZ|^{1/2}}\exp\left\{-\frac{1}{2 \hat{\sigma}^2_{\epsilon}}\|\uP_{\uZ}(\uy-\uR_{\omega}\utheta_{\omega})\|_2^2 \right\},
\end{eqnarray}
where $\hat{\sigma}^2_{\epsilon}=n^{-1} \|\uy-\hat{\uy}_\omega \|_2^2$. To complete our Bayesian model specification, we need to define prior distributions for unknown parameters. Given $\omega$, we assign a flat prior for $\utheta_\omega$, $\pi(\utheta_\omega \mid \omega)\propto 1$, so that our posterior inference is comparable to conventional generalized method of moments estimation. In a Bayesian framework, the model uncertainty due to the unknown invalid instruments can be addressed by treating $\omega$ as a nuisance parameter and averaging over $\omega$ in the joint posterior inference. \citet{kang2016instrumental} establish the identifiability condition of invalid instrumental variables such that $|\omega^*|< p/2$, where $\omega^*$ is the index set of the true invalid instruments and $|\cdot|$ denotes the cardinality of a set. To reflect this prior knowledge in our posterior inference, we use $\pi(\omega)\propto 1(|\omega|< p/2)$, where $1(\cdot)$ is an indicator function. Since we use the improper uniform prior for $\utheta_\omega$ given $\omega$, it is necessary to verify the propriety of the posterior distribution. The following theorem shows that our posterior distribution is proper under mild conditions.
\begin{theorem}\label{thm:0}
If $\uZ$ has full column rank and $\uR_{\omega}$ has full column rank for any $\omega$ such that $|\omega|<p/2$, then the posterior distribution is proper, that is,
$$ \sum_{\omega} \int \tilde{f}( \uy \mid \utheta_{\omega}, \omega)  \pi(\utheta_\omega \mid \omega)   \pi(\omega)  d\utheta_\omega  <\infty.$$
\end{theorem}
The proof of Theorem \ref{thm:0} is given in Appendix \ref{App:0}.

We now describe how to implement the posterior inference for the causal effect, $\beta$. Let $\mathcal{D}=(\uy,\ud,\uZ)$ be the observed data. The traditional Bayesian inference for $\beta$ can be made by using the marginal posterior distribution of $\beta$,
\begin{eqnarray}\label{Bayes:est11}
\pi(\beta\mid \mathcal{D})&=&  \sum_{\omega \in \Omega} \pi (\beta\mid \omega, \mathcal{D}) \pi(\omega\mid \mathcal{D}),
\end{eqnarray}
where $\Omega=\{\omega:|\omega|< p/2\}$. However, \citet{madigan1994model} argue that averaging over all possible candidate models as in \eqref{Bayes:est11} does not accurately represent the true model uncertainty if some models are substantially far away from the true model. As an alternative, they propose to exclude the models that predict the data far less well than the best predictive model from the posterior inference. Thus, we define a set of the models that provides a comparable performance to that of the best predictive model as
\begin{eqnarray}\label{compare:1}
\mathcal{A}=\{\omega \in \Omega : \max_{\omega' \in \Omega} \pi(\omega'\mid \mathcal{D})/\pi(\omega\mid \mathcal{D})\leq  c \}
\end{eqnarray}
for a pre-specified value of $c \geq 1$. If $c=1$, then $\mathcal{A}$ contains only the best predictive model and if $c=\infty$, then $\mathcal{A}$ consists of all the candidate models, i.e., $\mathcal{A}=\Omega$. In this paper, we set $c=3$ so that the evidence against any model in $\mathcal{A}$ (comparing to the highest posterior probability model) is not worth more than a bare mention \citep{kass1995bayes}. 

Given the truncated model space $\mathcal{A}$, the posterior distribution of $\beta$ is then defined as
\begin{eqnarray}\label{Bayes:est2}
\pi_{\mathcal{A}}(\beta\mid \mathcal{D})=  \sum_{\omega \in  \mathcal{A} } \pi (\beta\mid \omega, \mathcal{D}) \pi_{\mathcal{A}}(\omega\mid \mathcal{D}),
\end{eqnarray}
where $\pi_{\mathcal{A}}(\omega\mid \mathcal{D})=\pi(\omega\mid \mathcal{D})/\{\sum_{\omega' \in  \mathcal{A} }\pi(\omega'\mid \mathcal{D})\}$. In this paper, we propose to perform a Bayesian casual inference based on the posterior distribution of $\eqref{Bayes:est2}$.

Under our Bayesian framework, it is straightforward to derive an explicit form of $\pi(\beta\mid \omega, \mathcal{D})$. Since $\pi(\utheta_\omega \mid \omega)\propto 1$, by Bayes' theorem, deriving a quadratic form of $\theta_\omega$ from the logarithm of the pseudo-likelihood \eqref{like:1} immediately leads to
\begin{eqnarray*}\label{fullcon:1}
 \utheta_\omega \mid  \omega,\mathcal{D} \sim \mathcal{N}\{(\uR_{\omega}^\T \uP_{\uZ}\uR_{\omega} )^{-1} \uR_{\omega}^\T  \uP_{\uZ}\uy,  \hat{\sigma}^2_{\epsilon} (\uR_{\omega}^\T \uP_{\uZ} \uR_{\omega})^{-1} \},
 \end{eqnarray*} which implies that
\begin{eqnarray}\label{fullcon:1-1}
 \beta \mid \omega, \mathcal{D} \sim \mathcal{N}\{(\hat{\ud}_{\uZ}^{\T } \uM_{\uZ_\omega} \hat{\ud}_{\uZ} )^{-1} \hat{\ud}_{\uZ}^{\T} \uM_{\uZ_\omega}  \uy,  \hat{\sigma}^2_{\epsilon} (\hat{\ud}_{\uZ}^{\T} \uM_{\uZ_\omega} \hat{\ud}_{\uZ} )^{-1}  \},
 \end{eqnarray} 
 where $\uM_{\uZ_{\omega}}= \uI_n- \uZ_{\omega}(\uZ_{\omega}^\T \uZ_{\omega})^{-1} \uZ_{\omega}^\T $. Similarly, it is straightforward to show that the marginal pseudo-likelihood is 
 \begin{eqnarray*}\label{thm:fact1}
\tilde{f}(\uy\mid\omega)&=& \int  \tilde{f}( \uy \mid \utheta_{\omega}, \omega )\pi(\utheta_\omega \mid  \omega)  d \utheta_{\omega} \\
&\propto&  \frac{(2\pi \hat{\sigma}^2_{\epsilon})^{(|\omega|+1)/2} }{ |\uR_{\omega}^\T \uP_{\uZ} \uR_{\omega} |^{1/2}}\exp\left\{-\frac{1}{2 \hat{\sigma}^2_{\epsilon}}\|\uP_{\uZ}(\uy-\hat{\uy}_\omega )\|_2^2 \right\}.
 \end{eqnarray*} Under our prior specification, it follows from Bayes' theorem that $
 \pi(\omega \mid  \mathcal{D})\propto \tilde{f}( \uy \mid \omega )$ for $\omega \in \Omega$. Hence, calculating $\pi_{\mathcal{A}}(\omega\mid \mathcal{D})$ in \eqref{Bayes:est2} is straightforward as soon as $\mathcal{A}$ is obtained. 
 
However, in practice, constructing $\mathcal{A}$ is often computationally too expansive or infeasible. To overcome computational difficulties associated with constructing $\mathcal{A}$, we propose a new stochastic search algorithm that is motivated by the shotgun stochastic search algorithm of \citet{hans2007shotgun} but stimulates more dynamic movement of stochastic search via the notion of escort distribution used in thermodynamics. The concept of escort distribution is originally introduced by \citet{beck1995thermodynamics} for the characterization of chaos and multifractals. Let $p(x)$ be a probability mass function with support $\mathcal{X}$. An escort distribution of $p(x)$ of order $\tau$ is obtained by a deformation of the original distribution as follows:
 $$ p_\tau(x)=\frac{\{p(x)\}^\tau}{\sum_{x\in \mathcal{X}} \{p(x)\}^\tau },\quad \tau>0.$$
As $\tau \to 0$, the escort distribution converges to the uniform distribution. As $\tau \to \infty$, the escort distribution converges to the point mass distribution at the mode of $p(x)$.

A key idea of our stochastic search algorithm is to calculate $\mathcal{A}$ by constructing a Markov chain with escort distributions of $\pi(\omega \mid \mathcal{D})$ as follows: Set an initial model $\omega^{(1)}$ and $\mathcal{A}=\{\omega^{(1)}\}$. Repeat the following three steps for $t=1,\ldots,T$:
\begin{itemize}
\item[]\textbf{Step 1}: Compute a neighborhood of $\omega^{(t)}$ as $${\rm nbd}(\omega^{(t)})= \{\omega^{(t)} \setminus \{j\}:j\in \omega^{(t)}\} \cup \{ \omega^{(t)} \cup \{j'\}:j'\notin\omega^{(t)}\}.$$

\item[]\textbf{Step 2}: Define $\mathcal{W}= \mathcal{A} \cup {\rm nbd}(\omega^{(t)})$ and then update $$\mathcal{A}=\{\omega \in \mathcal{W} : \max_{\omega' \in \mathcal{W}} \pi(\omega'\mid \mathcal{D})/\pi(\omega\mid \mathcal{D})\leq  c \}.$$

\item[]\textbf{Step 3}: Update $\omega^{(t+1)}$ by selecting a sample from ${\rm nbd}(\omega^{(t)})$ with probabilities 
\begin{eqnarray}\label{escort}
 \pi_{\tau}(\omega \mid \mathcal{D})=\frac{\{\pi(\omega \mid \mathcal{D})\}^\tau}{\sum_{\omega' \in {\rm nbd}(\omega^{(t)})} \{\pi(\omega' \mid \mathcal{D}) \}^\tau }.
 \end{eqnarray}
\end{itemize}
After running the proposed algorithm until convergence, the set $\mathcal{A}$ contains only the acceptable models defined in \eqref{compare:1}. 
In the proposed stochastic search algorithm, by choosing $\tau \in (0,1)$, the escort distribution \eqref{escort} reduces the chance of getting stuck in a local maximum. For the simulation study and the real data application that are illustrated in the later sections, we set $\tau=0.1$ so that a substantial difference between two original marginal likelihoods is adjusted to almost no difference between their escort distributions according to the Jeffreys guideline \citep{jeffreys1998theory}.

The following theorem shows an important asymptotic property of the proposed method.
\begin{theorem}\label{thm:1}
Let $\omega^*$ be the index set of the true invalid instruments in the outcome model \eqref{m:1}. Under the regularity conditions of \citet{hansen1982large} that leads to \eqref{lim:dis}, our Bayesian model selection is consistent in the sense that
$$\pi(\omega^*\mid \mathcal{D})\to 1$$
in probability as $n\to \infty$.
\end{theorem}
The proof of Theorem \ref{thm:1} is given in Appendix \ref{App:1}. From the definition of the posterior distribution in \eqref{Bayes:est2}, Theorem  \ref{thm:1} implies that $\pi_{\mathcal{A}}(\beta\mid \mathcal{D})\approx\pi(\beta\mid \omega^*, \mathcal{D})$ for sufficiently large $n$. Hence, it follows from the result in \eqref{fullcon:1-1} that our posterior inference is asymptotically equivalent to the limiting distribution of the oracle two-stage least squares  estimator that is obtained by \eqref{oracle:normality} for a given $\omega^*$. In other words, when the sample size is large, the proposed method is as good as the ideal case, which is the oracle estimator with the known set of true valid instruments.

\section{Simulation study}\label{sec:4}
We now conduct a simulation study to examine the finite-sample performance of the proposed Bayesian estimator. We consider two different data-generating models. For model $1$, data are independently simulated from the following hierarchical model:
\begin{eqnarray*}
y_i=\beta d_i+\sum_{j=1}^{12} \alpha_j z_{ij}+\epsilon_i, \quad 
d_i= \sum_{j=1}^{12}  \eta_j z_{ij}+\nu_i,
\end{eqnarray*}
where we set $\alpha_1=\alpha_2=\alpha_3=0.5$ and $\alpha_4=\cdots\alpha_{12}=0$, varying values are considered for $\beta$ and $\eta_j$, $z_{ij}$ are independently generated from $\mathcal{N}(0,1)$, and the errors are generated from the bivariate normal distribution as follows:
\begin{eqnarray}\label{error:1}
\left(\begin{matrix}
\epsilon_i \\ \nu_i \end{matrix} \right)
 \sim \mathcal{N}\left\{\left(\begin{matrix}
0 \\ 0 
\end{matrix}\right), \left(\begin{matrix}
1 & 0.25  \\ 0.25 & 1   
\end{matrix} \right) \right\},
\end{eqnarray}
where non-zero off-diagonals represent correlations between $d_i$ and $\epsilon_i$.

For model $2$, we consider the same setup as in model $1$ except that the errors are generated from the asymmetric bivariate Laplace distribution using the following representation \citep{kozubowski2001asymmetric}: $(\epsilon_i , \nu_i)^\T =\sqrt{v_i}\ue_i$, where $v_i$ follows the exponential distribution with mean $1$ and $\ue_i=(e_{i1},e_{i2})^\T$ follows the bivariate normal distribution defined in \eqref{error:1}. For both models, the first three instruments are invalid and the remaining instruments are valid.

Under each model, we consider four different settings for $\beta$ and $\eta_j$: 
\begin{itemize}
\item[(a)] $\beta=0$, $\eta_{1}=\cdots=\eta_{12}=0.4$; 
\item[(b)] $\beta=0$, $\eta_1=\eta_2=\eta_3=0.6$, $\eta_{4}=\cdots=\eta_{12}=0.2$; 
\item[(c)] $\beta=0.5$, $\eta_{1}=\cdots=\eta_{12}=0.4$; 
\item[(d)] $\beta=0.5$, $\eta_1=\eta_2=\eta_3=0.6$, $\eta_{4}=\cdots=\eta_{12}=0.2$. 
\end{itemize}
In cases (a) and (b), the exposure has no effect on the outcome. In cases (b) and (d), the valid instrumental variables have relatively weaker correlations with the exposure than the invalid instrumental variables.
For each case, we consider two different sample sizes, $n=500$ and $n=2,000$.

Here, the parameter of interest is $\beta$. For the proposed Bayes method, after running the proposed stochastic search algorithm with $1,000$ iterations, posterior mean, posterior variance and 95\% credible interval are obtained from the marginal posterior distribution defined in \eqref{Bayes:est2}. To demonstrate the validity of the proposed Bayesian inference, the traditional Bayesian model averaging method is also performed by using the conventional posterior distribution given in \eqref{Bayes:est11}. In addition, the following seven existing methods are employed for comparison: (i) the na\"ive two-stage least squares (TSLS) estimator that treats all instruments as valid; (ii) the oracle TSLS estimator that uses the true valid instruments; (iii) the median estimator proposed by \citet{han2008detecting}; (iv) the lasso estimator proposed by \citet{kang2016instrumental}; (v) the adaptive lasso estimator proposed by \citet{windmeijer2018use}; (vi) the post-lasso estimator that is the TSLS estimator with the valid instruments selected by the lasso; and (vii) the post-adaptive lasso estimator that is the TSLS estimator with the valid instruments selected by the adaptive lasso. In our simulation study, the lasso and the adaptive lasso methods are implemented by the \texttt{R} package  \texttt{glmnet} with 10-fold cross-validation. The oracle TSLS estimator serves as a benchmark.

Table \ref{table:0} summarizes the simulation results with $n=500$ over $3,000$ Monte Carlo experiments. For each scenario, the proposed Bayes estimator outperforms all the existing estimators, except the oracle TSLS estimator. Our results clearly show that the proposed estimator has minimum levels of bias and good coverage rates of $95\%$ credible intervals for both Gaussian and non-Gaussian errors. In addition, the performance of the proposed Bayes estimator is always comparable to that of the oracle TSLS estimator, and this is consistent with our theoretical results discussed in Section \ref{sec:3}. 

Although the traditional Bayes estimator tends to have slightly smaller mean squared error than the proposed method, it produces larger variances than the proposed Bayes estimator and, as a result, leads to high inflation on coverage rates for $95\%$ credible intervals. This indicates that averaging over all possible models leads to overestimation of the model uncertainty. In contrast, the post-lasso estimator and the post-adaptive lasso estimator provide relatively poor coverage rates of $95\%$ confidence intervals due to the fact that penalized or post-penalized estimation methods ignore the uncertainty associated with instrumental variable selection. 

Interestingly but not surprisingly, both Bayes estimators provide smaller bias than the oracle TSLS estimator. This is due to fact that the TSLS estimator displays greater bias as the number of instrumental variables used in the estimator increases when instruments are weakly correlated with the exposure \citep{phillips1980exact, buse1992bias}. While the oracle estimator uses all the valid instruments, Bayes estimators incorporate the cases with less numbers of valid instruments which leads to the bias reduction in the Bayes estimators. The simulation results with $n=2,000$ are shown in Table  \ref{table:1}. The results are similar to and consistent with the results with $n=500$.  

\begin{table}
\caption{Simulation results with sample size $n=500$ over $3,000$ Monte Carlo experiments. Abbreviations: $95\%$ CP, coverage probability of $95\%$ confidence interval or credible interval; MSE, mean squared error; N/A, not available; TSLS, two-stage least squares; Var, average of variance estimate. \label{table:0} }
\centering
\fbox{
\begin{tabular}{l|cccc|cccc}
& \multicolumn{4}{c|}{Model 1}&\multicolumn{4}{c}{Model 2}\\
 Method &  Bias & Var & MSE & $95\%$ CP &Bias & Var & MSE & $95\%$ CP \\
\hline
& \multicolumn{4}{c|}{Case (a)}& \multicolumn{4}{c}{Case (a)}\\
na\"ive TSLS  & 0.3116 & 0.0016 & 0.0990 & 0.0000& 0.3111 & 0.0016 & 0.0987 & 0.0000  \\ 
median & 0.0473 & N/A & 0.0042 & N/A & 0.0471 & N/A & 0.0042 &N/A \\ 
lasso  & 0.0498 & N/A & 0.0045 & N/A& 0.0498 & N/A & 0.0045 & N/A   \\ 
post-lasso   & 0.0598 & 0.0063 & 0.0084 & 0.9293 & 0.0600 & 0.0062 & 0.0085 & 0.9250  \\ 
adaptive lasso  & 0.0297 & N/A & 0.0032 & N/A & 0.0300 & N/A & 0.0033 & N/A \\ 
post-adaptive lasso & 0.0355 & 0.0039 & 0.0040 & 0.9220 & 0.0363 & 0.0039 & 0.0043 & 0.9253 \\ 
traditional Bayes  & 0.0012 & 0.0020 & 0.0015 & 0.9730 & 0.0007 & 0.0020 & 0.0015 & 0.9760  \\ 
proposed Bayes   & 0.0010 & 0.0017 & 0.0017 & 0.9463& 0.0005 & 0.0017 & 0.0017 & 0.9523   \\ 
oracle TSLS  & 0.0027 & 0.0014 & 0.0014 & 0.9527 & 0.0021 & 0.0014 & 0.0014 & 0.9520  \\ 
  \hline
  & \multicolumn{4}{c|}{Case (b)}& \multicolumn{4}{c}{Case (b)}\\
na\"ive TSLS & 0.6200 & 0.0017 & 0.3863 & 0.0000 & 0.6194 & 0.0017 & 0.3855 & 0.0000   \\ 
median & 0.0930 & N/A & 0.0162 & N/A & 0.0935 & N/A & 0.0168 & N/A  \\ 
lasso & 0.3449 & N/A & 0.1460 & N/A  & 0.3489 & N/A & 0.1495 & N/A \\ 
post-lasso& 0.3578 & 0.0397 & 0.1588 & 0.6510& 0.3602 & 0.0409 & 0.1612 & 0.6527   \\ 
adaptive lasso& 0.0722 & N/A & 0.0146 & N/A  & 0.0733 & N/A & 0.0155 & N/A \\ 
post-adaptive lasso& 0.0792 & 0.0208 & 0.0188 & 0.9410 & 0.0816 & 0.0213 & 0.0203 & 0.9330  \\ 
traditional Bayes   &0.0058& 0.0083 & 0.0065 & 0.9787 & 0.0043 & 0.0083 & 0.0063 & 0.9803 \\ 
proposed Bayes & 0.0042 & 0.0068 & 0.0072 & 0.9483 & 0.0027 & 0.0068 & 0.0070 & 0.9520  \\ 
oracle TSLS & 0.0101 & 0.0054 & 0.0055 & 0.9490 & 0.0087 & 0.0054 & 0.0054 & 0.9513 \\ 
    \hline
      & \multicolumn{4}{c|}{Case (c)}& \multicolumn{4}{c}{Case (c)}\\
na\"ive TSLS & 0.3122 & 0.0016 & 0.0994 & 0.0000 & 0.3122 & 0.0016 & 0.0993 & 0.0000  \\ 
median & 0.0469 & N/A & 0.0042 & N/A & 0.0467 &N/A & 0.0042 & N/A \\ 
lasso & 0.0496 & N/A & 0.0045 & N/A & 0.0492 & N/A & 0.0045 & N/A\\ 
post-lasso & 0.0600 & 0.0061 & 0.0084 & 0.9320& 0.0586 & 0.0062 & 0.0081 & 0.9307  \\ 
adaptive lasso & 0.0297 & N/A & 0.0033 & N/A & 0.0291 & N/A & 0.0032 & N/A \\ 
post-adaptive lasso & 0.0359 & 0.0038 & 0.0041 & 0.9273 & 0.0349 & 0.0038 & 0.0040 & 0.9277 \\ 
traditional Bayes & 0.0009 & 0.0020 & 0.0016 & 0.9740 & 0.0012 & 0.0020 & 0.0015 & 0.9723 \\ 
proposed Bayes & 0.0006 & 0.0017 & 0.0017 & 0.9500 & 0.0008 & 0.0017 & 0.0017 & 0.9483 \\ 
oracle TSLS & 0.0022 & 0.0014 & 0.0014 & 0.9460  & 0.0028 & 0.0014 & 0.0014 & 0.9497 \\ 
  \hline
    & \multicolumn{4}{c|}{Case (d)}& \multicolumn{4}{c}{Case (d)}\\
na\"ive TSLS & 0.6204 & 0.0017 & 0.3868 & 0.0000 & 0.6194 & 0.0017 & 0.3855 & 0.0000 \\ 
median & 0.0927 & N/A & 0.0167 & N/A & 0.0939 & N/A & 0.0169 & N/A \\ 
lasso & 0.3445 & N/A & 0.1471 & N/A& 0.3497 & N/A & 0.1510 & N/A \\ 
post-lasso  & 0.3570 & 0.0396 & 0.1596 & 0.6533 & 0.3623 & 0.0407 & 0.1632 & 0.6410 \\ 
adaptive lasso  & 0.0723 & N/A & 0.0152 & N/A & 0.0728 & N/A & 0.0153 & N/A \\ 
post-adaptive lasso  & 0.0806 & 0.0207 & 0.0202 & 0.9423 & 0.0789 & 0.0208 & 0.0186 & 0.9360 \\ 
traditional Bayes & 0.0049 & 0.0083 & 0.0066 & 0.9743 & 0.0045 & 0.0083 & 0.0064 & 0.9733 \\ 
proposed Bayes & 0.0032 & 0.0068 & 0.0072 & 0.9493& 0.0022 & 0.0068 & 0.0069 & 0.9453 \\ 
oracle TSLS & 0.0091 & 0.0054 & 0.0057 & 0.9467  & 0.0096 & 0.0054 & 0.0055 & 0.9457\\ 
\end{tabular}}
\end{table}

\begin{table}
\caption{Simulation results with sample size $n=2,000$ over $3,000$ Monte Carlo experiments. Abbreviations: $95\%$ CP, coverage probability of $95\%$ confidence interval or credible interval; MSE, mean squared error; N/A, not available; TSLS, two-stage least squares; Var, average of variance estimate.\label{table:1}}
\centering
\fbox{
\begin{tabular}{l|cccc|cccc}
& \multicolumn{4}{c|}{Model 1}&\multicolumn{4}{c}{Model 2}\\
 Method &  Bias & Var & MSE & $95\%$ CP &Bias & Var & MSE & $95\%$ CP \\
\hline
  & \multicolumn{4}{c|}{Case (a)}& \multicolumn{4}{c}{Case (a)}\\
na\"ive TSLS  & 0.3121 & 0.0004 & 0.0979 & 0.0000 & 0.3125 & 0.0004 & 0.0981 & 0.0000 \\
median & 0.0233 & N/A  & 0.0010 & N/A  & 0.0241 & N/A  & 0.0011 &N/A  \\  
lasso& 0.0244 & N/A  & 0.0011 & N/A & 0.0253 & N/A  & 0.0011 & N/A  \\  
post-lasso & 0.0307 & 0.0011 & 0.0036 & 0.9133& 0.0336 & 0.0011 & 0.0043 & 0.9050 \\ 
adaptive lasso & 0.0097 &N/A  & 0.0006 & N/A   & 0.0105 & N/A  & 0.0006 &N/A  \\ 
post-adaptive lasso & 0.0121 & 0.0006 & 0.0009 & 0.8943  & 0.0127 & 0.0006 & 0.0009 & 0.8973 \\
traditional Bayes & 0.0003 & 0.0005 & 0.0004 & 0.9707 & 0.0006 & 0.0005 & 0.0004 & 0.9723 \\ 
proposed Bayes & 0.0003 & 0.0004 & 0.0004 & 0.9467 & 0.0007 & 0.0004 & 0.0004 & 0.9500 \\ 
oracle TSLS & 0.0006 & 0.0003 & 0.0003 & 0.9477 & 0.0009 & 0.0003 & 0.0003 & 0.9553 \\
\hline
  & \multicolumn{4}{c|}{Case (b)}& \multicolumn{4}{c}{Case (b)}\\
na\"ive TSLS & 0.6230 & 0.0004 & 0.3887 & 0.0000 & 0.6229 & 0.0004 & 0.3885 & 0.0000 \\ 
median & 0.0458 & N/A  & 0.0040 &N/A  & 0.0477 &N/A  & 0.0042 &N/A \\ 
lasso& 0.1619 & N/A  & 0.0304 & N/A  & 0.1630 & N/A  & 0.0307 &N/A  \\
post-lasso & 0.1627 & 0.0099 & 0.0325 & 0.6993 & 0.1648 & 0.0098 & 0.0335 & 0.6983 \\ 
adaptive lasso  & 0.0294 & N/A  & 0.0032 & N/A  & 0.0317 & N/A  & 0.0034 &N/A \\ 
post-adaptive lasso & 0.0348 & 0.0040 & 0.0043 & 0.9343 & 0.0378 & 0.0040 & 0.0048 & 0.9210 \\ 
traditional Bayes & 0.0001 & 0.0018 & 0.0015 & 0.9707 & 0.0020 & 0.0018 & 0.0015 & 0.9723 \\ 
proposed Bayes & 0.0011 & 0.0015 & 0.0021 & 0.9497& 0.0032 & 0.0015 & 0.0023 & 0.9477 \\ 
oracle TSLS & 0.0011 & 0.0014 & 0.0014 & 0.9493 & 0.0030 & 0.0014 & 0.0013 & 0.9553 \\ 
\hline
  & \multicolumn{4}{c|}{Case (c)}& \multicolumn{4}{c}{Case (c)}\\
na\"ive TSLS & 0.3123 & 0.0004 & 0.0980 & 0.0000 & 0.3126 & 0.0004 & 0.0981 & 0.0000\\ 
median & 0.0240 & N/A & 0.0011 &N/A & 0.0234 &N/A  & 0.0010 &N/A \\ 
 lasso & 0.0251 & N/A & 0.0011 & N/A  & 0.0246 &N/A  & 0.0011 &N/A  \\ 
post-lasso & 0.0313 & 0.0011 & 0.0035 & 0.9107& 0.0323 & 0.0011 & 0.0041 & 0.9137 \\ 
adaptive lasso  & 0.0101 & N/A & 0.0006 & N/A & 0.0097 & N/A & 0.0006 &N/A   \\ 
post-adaptive lasso & 0.0126 & 0.0006 & 0.0008 & 0.9000& 0.0120 & 0.0006 & 0.0009 & 0.9037  \\
traditional Bayes & 0.0007 & 0.0005 & 0.0004 & 0.9707 & 0.0003 & 0.0005 & 0.0004 & 0.9710  \\
proposed Bayes & 0.0007 & 0.0004 & 0.0004 & 0.9497& 0.0003 & 0.0004 & 0.0004 & 0.9543\\
oracle TSLS & 0.0007 & 0.0003 & 0.0003 & 0.9500 & 0.0004 & 0.0003 & 0.0003 & 0.9547 \\ 
\hline
  & \multicolumn{4}{c|}{Case (d)}& \multicolumn{4}{c}{Case (d)}\\
na\"ive TSLS & 0.6240 & 0.0004 & 0.3899 & 0.0000 & 0.6243 & 0.0004 & 0.3903 & 0.0000 \\  
median & 0.0475 & N/A  & 0.0042 &N/A  & 0.0456 & N/A & 0.0040 & N/A \\  
lasso & 0.1614 & N/A & 0.0304 & N/A & 0.1632 & N/A  & 0.0311 &N/A  \\ 
post-lasso & 0.1637 & 0.0097 & 0.0336 & 0.6803 & 0.1632 & 0.0100 & 0.0323 & 0.7053  \\ 
adaptive lasso  & 0.0316 &N/A  & 0.0033 &N/A & 0.0290 &N/A  & 0.0031 &N/A   \\ 
post-adaptive lasso & 0.0379 & 0.0039 & 0.0046 & 0.9250 & 0.0343 & 0.0040 & 0.0042 & 0.9347 \\ 
traditional Bayes & 0.0020 & 0.0018 & 0.0015 & 0.9713 & 0.0006 & 0.0018 & 0.0015 & 0.9730  \\ 
proposed Bayes & 0.0029 & 0.0015 & 0.0021 & 0.9493& 0.0019 & 0.0015 & 0.0025 & 0.9510  \\ 
oracle TSLS & 0.0027 & 0.0014 & 0.0014 & 0.9513 & 0.0020 & 0.0014 & 0.0014 & 0.9540  \\ 
\end{tabular}
}
\end{table}

\section{Real data application}\label{sec:5}
We apply the proposed Bayesian method to a classic example of many instruments: \citet{angrist1991does} on the impact of compulsory schooling on earnings. We use three subsets of data from the 1970 and 1980 Census. The first dataset contains information on $247,199$ men born between 1920 and 1929 in the 1970 Census. The second dataset consists of 329,509 men born between 1930 and 1939 in the 1980 Census. The third dataset includes 486,926 men born 1940 and 1949 from the 1980 Census. The outcome variable is the log of weekly earnings and the exposure variable is the years of schooling completed. 

In the original study of \citet{angrist1991does}, 30 quarter-of-birth dummies (3 quarters $\times$ 10 years) are used as instrumental variables. A number of papers examine whether the quarter-of-birth dummies are appropriate as the instruments for answering the question of how the years of schooling affect earnings \citep[e.g.][]{bound1995problems,imbens2005robust, buckles2013season,fan2017quarter}. While the majority of this literature, including the influential works of \citet{bound1995problems,imbens2005robust} focus on whether the results of \citet{angrist1991does} are driven by the weak association between the quarter-of-birth dummies and the schooling (i.e. the weak IV problem), several recent works such as \citet{buckles2013season,fan2017quarter} investigate the possibility of the quarter-of-brith dummies violating the exclusion restriction and result with mixed conclusions. We revisit the analysis of \citet{angrist1991does} to demonstrate our proposed method and to contribute to this recent development in the literature.

We consider the following outcome model:
\begin{eqnarray}
y_i=\beta d_i +\ualpha_1^{\T} \uz_{1i} +\ualpha_2^{\T} \uz_{2i}+\epsilon_i,
\end{eqnarray}
where $y_i$ is the log of weekly wage of the $i$th individual, $d_i$ is the years of education, $\uz_{1i}$ is the vector of quarter-of-birth dummies, and $\uz_{2i}$ is the vector of year-of-birth dummies. The parameter of interest, $\beta$, represents the effect of education on earnings, i.e., the return to education. Similar to the original study of \citet{angrist1991does}, we treat $\uz_{2i}$, the year-of-birth dummies as covariates, which is identical as treating them as the (known) invalid instruments by definition, whereas it is unknown whether some elements of $\uz_{1i}$, the quarter-of-birth dummies, are invalid or not. 

We estimate the model via the following estimators: i) a na\"ive estimator by treating all the quarter-of-birth dummies as valid instruments as in \citet{angrist1991does}; ii) the lasso estimator proposed by \citet{kang2016instrumental} along with the post-lasso estimator for the inference; iii) the adaptive lasso estimator proposed by \citet{windmeijer2018use} along with the post-adaptive lasso estimator for the inference; and iv) the proposed Bayesian estimator. To incorporate the treatment of $\uz_{2i}$, the year-of-birth dummies as covariates, the proposed Bayesian estimation is implemented by letting $\uz_i=(\uz_{1i}^{\T},\uz_{2i}^{\T})^{\T}$ and $\ualpha=(\ualpha_1^{\T},\ualpha_2^{\T})^\T$ and assuming that $\omega$ always includes $\{31,\ldots,39\}$ and similarly, the penalized least squares estimators are applied after removing $\ualpha_2$ from the penalty term in \eqref{pls}. 

Table \ref{table:real1} displays estimates of the return to education for the three data sets. For the 1920--1929 cohort, the na\"ive estimate of the return to education is 0.0769 with standard error of 0.0150. The proposed Bayes estimate of the return to education is 0.0794 with standard error of 0.0171 for men born 1920--1929 in the 1970 Census. It is important to note that the post-lasso estimate is obtained as 0.0695 with standard error of 0.0507 and the post-adaptive lasso estimate is 0.0715 with standard error of 0.0373. This implies that the penalized likelihood methods conclude that there is no return to education at 0.05 level of significance for men born 1920--1929 in the 1970 Census. For the 1930--1939 cohort, the na\"ive TSLS estimate and the Bayes estimate are the same as 0.0891 with standard error of 0.01610. The post-lasso estimate of the return to education is 0.0834 with standard error of 0.0398. The post-adaptive lasso method supports no education return by providing the estimate of 0.0972 with standard error of 0.0582. For the 1940--1949 cohort, the na\"ive estimate of the return to education is 0.0553 with standard error of 0.0138 and the Bayes estimate is 0.0566 with standard error 0.0138. The post-adaptive lasso estimate is 0.0900 with standard error of 0.0211. For the 1940--1949 cohort, the post-lasso estimate is not available due to the fact that the lasso method selects all the quarter-of-birth dummies as invalid instruments. Overall, the estimates from the proposed Bayesian estimator track the results of the original study whereas the penalized estimators deviate from them and the degree of the deviations varies by different cohorts. 

To further unpack the results of Table \ref{table:real1}, Table \ref{table:real2} summaries the percentage of selecting the quarter-of-birth dummy as a valid instrument for each method. For the proposed Bayes estimator, the marginal posterior probability is computed by summing the posterior model probabilities across models for each quarter-of-birth dummy. For the 1920--1929 cohort, our Bayes estimator concludes that the 10\textsuperscript{th} instrument has 50.7\% chance to be valid and the 23\textsuperscript{rd} instruments has 81.9\% chance to be valid while the remaining instruments are valid with certainty. For the 1930--1939 cohort, the proposed Bayes estimator selects all the instruments as being valid with certainty, which leads to the identical result as the na\"ive TSLS estimate in Table \ref{table:real1}. For the 1940--1949 cohort, the 4\textsuperscript{th} instrument is considered as being invalid with probability of one by the proposed Bayes estimator, while the 10\textsuperscript{th} and 24\textsuperscript{th} instruments have about 70\% chance to be valid and the others attain 100\% of being valid instruments. 

In general, the results from the proposed Bayes estimator support the results of \citet{angrist1991does} and suggest that the majority of the quarter-of-birth dummies are valid. We do, however, see the differences across cohorts in terms of which and how many instruments are invalid. This is consistent with the argument by \citet{fan2017quarter} that the validity of the quarter-of-birth dummies is context-dependent. It is also worth noting that the lasso and the adaptive lasso methods are too liberal for selecting invalid instruments. For all three cohorts, we observe that only handful of instruments are selected as valid instruments. Moreover, these penalized least squares estimators treat more than a half of the instruments as invalid instruments, which violates the identifiability condition of \citet{kang2016instrumental}.

\begin{table}
\caption{Estimates of return to education. Abbreviations: SE, standard error.\label{table:real1}}
\centering
\fbox{
\begin{tabular}{l|cc|cc|cc}
  &\multicolumn{2}{c|}{ Data 1 (1920--1929)} &\multicolumn{2}{c|}{ Data 2 (1930--1939)} &\multicolumn{2}{c}{ Data 3 (1940--1949)}\\
 Method & Estimate & SE& Estimate & SE& Estimate & SE \\ 
  \hline
na\"ive TSLS& 0.0769 & 0.0150  & 0.0891 & 0.0161  & 0.0553 & 0.0138\\ 
  lasso & 0.0685 &  N/A & 0.0855 & N/A& 0.0757 &  N/A\\ 
  post-lasso & 0.0695 & 0.0507& 0.0834 & 0.0398 & N/A & N/A \\ 
  adaptive lasso& 0.0715 & N/A & 0.0908 &  N/A & 0.0902 &  N/A \\ 
 post-adaptive lasso& 0.0715 & 0.0373& 0.0972 & 0.0582& 0.0900 & 0.0211 \\ 
  proposed Bayes& 0.0794 & 0.0171& 0.0891 & 0.0161 & 0.0566 & 0.0138 \\ 
\end{tabular}}
\end{table}

\begin{table}
\caption{Percentage of selecting a quarter-of-birth dummy as a valid instrument, where the marginal posterior probability is used for the proposed Bayesian method. Abbreviations: N, na\"ive TSLS; L, lasso; A, adaptive lasso; B, proposed Bayes.\label{table:real2} }
\centering
\fbox{
\begin{tabular}{c|rrrr|rrrr|rrrr}
  Predictor &\multicolumn{4}{c|}{ Data 1 (1920--1929)} &\multicolumn{4}{c|}{ Data 2 (1930--1939)} &\multicolumn{4}{c}{ Data 3 (1940--1949)}\\
number& N & L  & A &B  & N & L  & A &B  &N & L  & A &B  \\
  \hline
1 & 100 & 0 & 0 & 100   & 100 & 0 & 0 & 100  & 100 & 0 & 100 & 100   \\ 
2 &100& 0 & 100 & 100  &100& 0 & 0 & 100 & 100 & 0 & 0 & 100 \\ 
3 & 100& 0 & 0 & 100 & 100& 0 & 0 & 100& 100 & 0 & 100 & 100  \\ 
4 &100& 100 & 100 & 100& 100& 0 & 0 & 100& 100 & 0 & 0 & 0 \\ 
5&100& 0 & 0 & 100& 100& 100 & 0 & 100& 100 & 0 & 0 & 100 \\ 
6 &100& 0 & 0 & 100& 100& 0 & 100 & 100& 100 & 0 & 0 & 100 \\ 
7 &100& 0 & 0 & 100& 100& 0 & 0 & 100& 100 & 0 & 0 & 100 \\ 
8 &100& 0 & 0 & 100& 100& 0 & 0 & 100& 100 & 0 & 0 & 100 \\ 
9 &100& 0 & 0 & 100& 100& 0 & 0 & 100& 100 & 0 & 0 & 100 \\ 
10&100& 0 & 0 & 50.7& 100& 0 & 0 & 100& 100 & 0 & 0 & 70.1 \\ 
 11 &100& 0 & 0 & 100& 100& 100 & 0 & 100& 100 & 0 & 0 & 100 \\ 
 12& 100& 100 & 100 & 100& 100& 0 & 0 & 100& 100 & 0 & 0 & 100 \\ 
13&100 & 0 & 0 & 100& 100& 0 & 0 & 100& 100 & 0 & 0 & 100 \\ 
 14& 100& 0 & 100 & 100& 100& 0 & 0 & 100& 100 & 0 & 0 & 100 \\ 
 15 & 100& 0 & 100 & 100& 100& 0 & 0 & 100& 100 & 0 & 0 & 100 \\ 
16 & 100& 100 & 100 &100& 100& 0 & 0 & 100& 100 & 0 & 0 & 100 \\ 
17&100& 0 & 0 & 100& 100& 0 & 0 & 100& 100 & 0 & 0 & 100 \\ 
18 &100& 0 & 0 & 100& 100& 0 & 0 & 100& 100 & 0 & 0 & 100 \\ 
19 &100& 0 & 0 &100& 100& 0 & 0 & 100 & 100 & 0 & 0 & 100 \\ 
20 &100& 0 & 0 & 100& 100& 0 & 0 & 100& 100 & 0 & 0 & 100 \\ 
21 &100&0 & 0 & 100 & 100& 0 & 0 & 100& 100 & 0 & 0 & 100 \\ 
22 &100& 0 & 0 & 100 & 100& 0 & 0 & 100& 100 & 0 & 0 & 100 \\ 
23 &100& 0 & 0 & 81.9 & 100& 0 & 0 & 100& 100 & 0 & 0 & 100 \\ 
24&100& 0 & 0 & 100& 100& 0 & 0 & 100& 100 & 0 & 0 & 70.2 \\ 
25&100& 0 & 0 & 100& 100& 0 & 0 & 100& 100 & 0 & 0 & 100 \\ 
26&100& 0 & 0 & 100& 100& 0 & 0 & 100& 100 & 0 & 0 & 100 \\ 
27 & 100 & 0 & 100 & 100& 100& 0 & 0 & 100& 100 & 0 & 0 & 100 \\ 
28& 100 & 0 & 0 & 100& 100& 0 & 0 & 100& 100 & 0 & 0 & 100 \\ 
29 & 100 & 0 & 0 & 100& 100& 100 & 100 & 100& 100 & 0 & 0 & 100 \\ 
30 & 100 & 0 & 100 & 100& 100& 0 & 0 & 100& 100 & 0 & 0 & 100 \\ 
\end{tabular}}
\end{table}

\section{Concluding remarks}\label{sec:6}
The IV estimation has been a valuable tool for numerous observational studies in many disciplines. With better data availability, the uses of many IVs become more common. We have proposed a novel Bayesian estimator to consistently estimate the causal effects when there are possible invalid instruments. We show that this novel Bayesian estimator performs well in the simulation study and also apply our estimator to the well-known example of \citet{angrist1991does}. 

To conclude, we acknowledge that the proposed Bayesian approach relies on the conditional homoscedasticity, $E(\epsilon^2|Z)=\sigma^2_\epsilon$. Under a general form of conditional heteroscedasticity, the proposed method can be extended by modifying the pseudo-likelihood defined in \eqref{like:1} as follows:
$$
\tilde{f}( \uy \mid \utheta_{\omega}, \omega)=\frac{1}{(2\pi )^{p/2} |\hat{\uSigma}_m |^{1/2}}\exp\left\{-\frac{1}{2n}  (\uy-\uR_{\omega}\utheta_{\omega})^\T \uZ \hat{\uSigma}_m^{-1} \uZ^\T (\uy-\uR_{\omega}\utheta_{\omega}) \right\}, $$ 
where $\hat{\uSigma}_m= n^{-1} \uZ^\T {\rm Diag}\{(\uy-\hat{\uy}_{\omega})  (\uy-\hat{\uy}_{\omega})^\T\} \uZ $. Under mild regularity conditions, the model selection consistency of the extended method can be shown by a similar proof of Theorem \ref{thm:1}. As a result, the resulting estimator of the extended method will be asymptotically equivalent to the oracle generalized method of moment estimator obtained by the true valid instrumental variables.

\appendix
\section{Proof of Theorem \ref{thm:0}}\label{App:0}
Since $\uZ$ and $\uR_\omega$ are of full column rank, $\uP_{\uZ}\uR_\omega$ also has full column rank. Hence, $ \uR_\omega^\T \uP_{\uZ}\uR_\omega$ is invertible and $\hat{\utheta}_\omega=(\uR_{\omega}^\T \uP_{\uZ} \uR_{\omega} )^{-1} \uR_{\omega}^\T  \uP_{\uZ} \uy$ is well defined for each $\omega \in \{ \omega: |\omega|<p/2\}$. It is straightforward to show that
\begin{equation*} \|\uP_{\uZ}(\uy-\uR_{\omega}\utheta_{\omega} )\|_2^2=\|\uP_{\uZ}(\uy-\uR_{\omega}\hat{\utheta}_{\omega} )\|_2^2+\|\uP_{\uZ} \uR_{\omega}(\hat{\utheta}_{\omega}-\utheta_{\omega} )\|_2^2 ,
\end{equation*}
which is known as the Pythagorean theorem. This result leads to
\begin{eqnarray*}
\int \tilde{f}( \uy \mid \utheta_{\omega}, \omega)  \pi(\utheta_\omega \mid \omega) d\utheta_\omega =\frac{\exp\left\{-\frac{1}{2 \hat{\sigma}^2_{\epsilon}}\|\uP_{\uZ}(\uy-\uR_{\omega}\hat{\utheta}_{\omega})\|_2^2 \right\}}{(2\pi \hat{\sigma}^2_{\epsilon})^{(p-|\omega|-1)/2} |n^{-1} \uZ^\T \uZ|^{1/2}|\uR_{\omega}^\T \uP_{\uZ} \uR_{\omega}|^{1/2}}<\infty,
\end{eqnarray*}
for any $\omega  \in \{ \omega: |\omega|<p/2\}$. Hence, we have 
$$\sum_{\omega} \int \tilde{f}( \uy \mid \utheta_{\omega}, \omega)  \pi(\utheta_\omega \mid \omega)   \pi(\omega)  d\utheta_\omega  <  \max_{\omega:|\omega|<p/2}\int \tilde{f}( \uy \mid \utheta_{\omega}, \omega)  \pi(\utheta_\omega \mid \omega) d\utheta_\omega <\infty.$$
\hfill $\square$

\section{Proof of Theorem \ref{thm:1}}\label{App:1}
Define $\Omega_0=\Omega\setminus \{ \omega^*\}$, where $\Omega=\{\omega: |\omega|< p/2\}$. Since $
 \pi(\omega \mid  \mathcal{D})\propto \tilde{f}( \uy \mid \omega )$ for $\omega \in \Omega$, 
$$ \pi(\omega^*\mid \mathcal{D}) =\frac{1}{1+\sum_{\omega \in \Omega_0}  \tilde{f}( \uy \mid \omega )/ \tilde{f}( \uy \mid \omega^* ) }.$$
Hence, it suffices to show that
$$  \tilde{f}( \uy \mid \omega )/ \tilde{f}(\uy \mid \omega^* ) \to 0 $$
in probability as $n\to \infty$ for any $\omega \in \Omega_0$. 

Under the regularity conditions, we have 
\begin{eqnarray}\label{lim:dis2}
\sqrt{n} m_n(\utheta_{\omega})\mid \utheta_{\omega},  \omega \to \mathcal{N}(0,\uSigma_m),
\end{eqnarray}
in distribution as $n\to \infty$ for $\omega ( \supset \omega^*)$, where $m_n(\utheta_{\omega})=n^{-1} \uZ^\T (\uy-\uR_{\omega}\utheta_{\omega})$ and $\uSigma_m=\sigma^2_\epsilon E(ZZ^\T)$. By the continuous mapping theorem, \eqref{lim:dis2} implies
$$\frac{1}{2 \hat{\sigma}^2_{\epsilon}}\|\uP_{\uZ}(\uy-\hat{\uy}_\omega )\|_2^2 \to \chi^2_{|\omega|}$$
in distribution as $n\to \infty$ for $\omega \supset \omega^*$, where $\chi^2_{|\omega|}$ denotes the chi-squared distribution with $|\omega|$ degrees of freedom. Note that
$$ n^{-1} \uR_{\omega}^\T \uP_{\uZ} \uR_{\omega} \to \left[\begin{matrix} E(D^2) & E(D Z_{\omega}^\T) \\ E(D Z_{\omega}) & E(Z_{\omega}Z_{\omega}^\T ) \end{matrix} \right] $$
in probability as $n\to \infty$. This implies that $| \uR_{\omega}^\T \uP_{\uZ} \uR_{\omega} |=O_p(n^{|\omega|})$.

First, assume $\omega$ to be an over-fitted model such that $\omega  \supset \omega^*$ and $\omega \neq \omega^*$. Recall that 
 \begin{eqnarray}\label{thm:fact1}
\tilde{f}(\uy\mid\omega) \propto  \frac{(2\pi \hat{\sigma}^2_{\epsilon})^{(|\omega|+1)/2} }{ |\uR_{\omega}^\T \uP_{\uZ} \uR_{\omega} |^{1/2}}\exp\left\{-\frac{1}{2 \hat{\sigma}^2_{\epsilon}}\|\uP_{\uZ}(\uy-\hat{\uy}_\omega )\|_2^2 \right\}.
 \end{eqnarray}
Since $\chi^2_{|\omega|}/\log n = o_p(1)$, it follows from \eqref{thm:fact1} that
$$2\log  \tilde{f}( \uy \mid \omega )-2\log \tilde{f}( \uy \mid \omega^* ) = (|\omega^*|-|\omega|) \log (n)\{1+o_p(1)\}.$$
As $|\omega|>|\omega^*|$, this implies that $\tilde{f}(\uy \mid \omega )/\tilde{f}( \uy \mid \omega^* )\to 0$ in probability as $n\to \infty$. 

Second, assume $\omega$ to be a mis-specified model such that $\omega \not\supset \omega^*$. Define $\omega^{\dagger}= \omega^* \cup \omega $. From the Pythagorean theorem, we have
\begin{equation}\label{A:1} \|\uP_{\uZ}(\uy-\uR_{\omega^{\dagger}}\utheta_{\omega^{\dagger}} )\|_2^2=\|\uP_{\uZ}(\uy-\uR_{\omega^{\dagger}}\hat{\utheta}_{\omega^\dagger} )\|_2^2+\|\uP_{\uZ} \uR_{\omega^{\dagger}}(\hat{\utheta}_{\omega^{\dagger}}-\utheta_{\omega^{\dagger}} )\|_2^2 ,
\end{equation}
where $\hat{\utheta}_{\omega^\dagger}=(\uR_{\omega^{\dagger}}^\T \uP_{\uZ} \uR_{\omega^{\dagger}} )^{-1} \uR_{\omega^{\dagger}}^\T  \uP_{\uZ} \uy$. Without loss of generality, suppose that $\utheta_{\omega^{\dagger}}=(\utheta_{\omega}^\T,\utheta_{\omega^{\dagger}\setminus \omega}^\T)^\T$. Letting $\utheta_{\omega^{\dagger}}=(\hat{\utheta}_{\omega}^\T,\0)^\T$ with $\hat{\utheta}_{\omega}=(\uR_{\omega}^\T \uP_{\uZ} \uR_{\omega} )^{-1} \uR_{\omega}^\T  \uP_{\uz} \uy$, Equation \eqref{A:1} reduces to 
$$ \|\uP_{\uZ}(\uy-\hat{\uy}_\omega  )\|_2^2=\|\uP_{\uZ}(\uy-\hat{\uy}_{\omega^\dagger} )\|_2^2+ n \tilde{\utheta}_{\omega}^\T (\tilde{\uS}_{\omega})^{-1} \tilde{\utheta}_{\omega},
 $$
where $\tilde{\uS}_{\omega}$ and $\tilde{\utheta}_{\omega}$ are the sub-matrix of $(n^{-1}\uR_{\omega^{\dagger}}^\T \uP_{\uZ} \uR_{\omega^{\dagger}})^{-1}$ and the sub-vector of $\hat{\utheta}_{\omega^{\dagger}}$ corresponding to $\omega$, respectively. Since $\tilde{\utheta}_{\omega}^\T (\tilde{\uS}_{\omega})^{-1} \tilde{\utheta}_{\omega}=O_p(1)$, it follows from \eqref{thm:fact1} that 
$$2\log \tilde{f}( \uy \mid \omega )-2\log \tilde{f}( \uy \mid \omega^* ) =  (|\omega^*|-|\omega|) \log (n)\{1+o_p(1)\}-O_p(n).$$
Hence, we have that $\tilde{f}( \uy \mid \omega )/\tilde{f}( \uy \mid \omega^* )\to 0$ in probability as $n\to \infty$ for any $\omega  \not\supset \omega^*$. This completes our proof. \hfill $\square$


\begin{thebibliography}{}

\bibitem[\protect\citeauthoryear{Andrews}{Andrews}{1999}]{andrews1999consistent}
Andrews, D.~W. (1999).
\newblock Consistent moment selection procedures for generalized method of
  moments estimation.
\newblock {\em Econometrica\/}~{\em 67\/}(3), 543--563.

\bibitem[\protect\citeauthoryear{Angrist, Imbens, and Rubin}{Angrist
  et~al.}{1996}]{angrist1996identification}
Angrist, J.~D., G.~W. Imbens, and D.~B. Rubin (1996).
\newblock Identification of causal effects using instrumental variables.
\newblock {\em Journal of the American Statistical Association\/}~{\em
  91\/}(434), 444--455.

\bibitem[\protect\citeauthoryear{Angrist and Krueger}{Angrist and
  Krueger}{1991}]{angrist1991does}
Angrist, J.~D. and A.~B. Krueger (1991).
\newblock Does compulsory school attendance affect schooling and earnings?
\newblock {\em The Quarterly Journal of Economics\/}~{\em 106\/}(4), 979--1014.

\bibitem[\protect\citeauthoryear{Angrist and Pischke}{Angrist and
  Pischke}{2010}]{angrist2010credibility}
Angrist, J.~D. and J.-S. Pischke (2010).
\newblock The credibility revolution in empirical economics: How better
  research design is taking the con out of econometrics.
\newblock {\em Journal of Economic Perspectives\/}~{\em 24\/}(2), 3--30.

\bibitem[\protect\citeauthoryear{Beck and Sch{\"o}gl}{Beck and
  Sch{\"o}gl}{1995}]{beck1995thermodynamics}
Beck, C. and F.~Sch{\"o}gl (1995).
\newblock {\em Thermodynamics of chaotic systems: an introduction}.
\newblock Number~4. Cambridge University Press.

\bibitem[\protect\citeauthoryear{Belloni, Chen, Chernozhukov, and
  Hansen}{Belloni et~al.}{2012}]{belloni2012sparse}
Belloni, A., D.~Chen, V.~Chernozhukov, and C.~Hansen (2012).
\newblock Sparse models and methods for optimal instruments with an application
  to eminent domain.
\newblock {\em Econometrica\/}~{\em 80\/}(6), 2369--2429.

\bibitem[\protect\citeauthoryear{Belloni, Chernozhukov, et~al.}{Belloni
  et~al.}{2013}]{belloni2013least}
Belloni, A., V.~Chernozhukov, et~al. (2013).
\newblock Least squares after model selection in high-dimensional sparse
  models.
\newblock {\em Bernoulli\/}~{\em 19\/}(2), 521--547.

\bibitem[\protect\citeauthoryear{Bound, Jaeger, and Baker}{Bound
  et~al.}{1995}]{bound1995problems}
Bound, J., D.~A. Jaeger, and R.~M. Baker (1995).
\newblock Problems with instrumental variables estimation when the correlation
  between the instruments and the endogenous explanatory variable is weak.
\newblock {\em Journal of the American statistical association\/}~{\em
  90\/}(430), 443--450.

\bibitem[\protect\citeauthoryear{Buckles and Hungerman}{Buckles and
  Hungerman}{2013}]{buckles2013season}
Buckles, K.~S. and D.~M. Hungerman (2013).
\newblock Season of birth and later outcomes: Old questions, new answers.
\newblock {\em Review of Economics and Statistics\/}~{\em 95\/}(3), 711--724.

\bibitem[\protect\citeauthoryear{Buse}{Buse}{1992}]{buse1992bias}
Buse, A. (1992).
\newblock The bias of instrumental variable estimators.
\newblock {\em Econometrica\/}~{\em 60\/}(1), 173--180.

\bibitem[\protect\citeauthoryear{Chernozhukov, Hansen, and
  Spindler}{Chernozhukov et~al.}{2015}]{chernozhukov2015post}
Chernozhukov, V., C.~Hansen, and M.~Spindler (2015).
\newblock Post-selection and post-regularization inference in linear models
  with many controls and instruments.
\newblock {\em American Economic Review\/}~{\em 105\/}(5), 486--90.

\bibitem[\protect\citeauthoryear{Chernozhukov and Hong}{Chernozhukov and
  Hong}{2003}]{chernozhukov2003mcmc}
Chernozhukov, V. and H.~Hong (2003).
\newblock An mcmc approach to classical estimation.
\newblock {\em Journal of Econometrics\/}~{\em 115\/}(2), 293--346.

\bibitem[\protect\citeauthoryear{Didelez and Sheehan}{Didelez and
  Sheehan}{2007}]{didelez2007mendelian}
Didelez, V. and N.~Sheehan (2007).
\newblock Mendelian randomization as an instrumental variable approach to
  causal inference.
\newblock {\em Statistical methods in medical research\/}~{\em 16\/}(4),
  309--330.

\bibitem[\protect\citeauthoryear{Fan, Liu, and Chen}{Fan
  et~al.}{2017}]{fan2017quarter}
Fan, E., J.-T. Liu, and Y.-C. Chen (2017).
\newblock Is the quarter of birth endogenous? new evidence from taiwan, the us,
  and indonesia.
\newblock {\em Oxford Bulletin of Economics and Statistics\/}~{\em 79\/}(6),
  1087--1124.

\bibitem[\protect\citeauthoryear{Han}{Han}{2008}]{han2008detecting}
Han, C. (2008).
\newblock Detecting invalid instruments using {L}1-{GMM}.
\newblock {\em Economics Letters\/}~{\em 101\/}(3), 285--287.

\bibitem[\protect\citeauthoryear{Hans, Dobra, and West}{Hans
  et~al.}{2007}]{hans2007shotgun}
Hans, C., A.~Dobra, and M.~West (2007).
\newblock Shotgun stochastic search for ``large p'' regression.
\newblock {\em Journal of the American Statistical Association\/}~{\em
  102\/}(478), 507--516.

\bibitem[\protect\citeauthoryear{Hansen}{Hansen}{1982}]{hansen1982large}
Hansen, L.~P. (1982).
\newblock Large sample properties of generalized method of moments estimators.
\newblock {\em Econometrica\/}~{\em 50}, 1029--1054.

\bibitem[\protect\citeauthoryear{Hoeting, Madigan, Raftery, and
  Volinsky}{Hoeting et~al.}{1999}]{hoeting1999bayesian}
Hoeting, J.~A., D.~Madigan, A.~E. Raftery, and C.~T. Volinsky (1999).
\newblock Bayesian model averaging: a tutorial.
\newblock {\em Statistical science\/}, 382--401.

\bibitem[\protect\citeauthoryear{Imbens and Rosenbaum}{Imbens and
  Rosenbaum}{2005}]{imbens2005robust}
Imbens, G.~W. and P.~R. Rosenbaum (2005).
\newblock Robust, accurate confidence intervals with a weak instrument: quarter
  of birth and education.
\newblock {\em Journal of the Royal Statistical Society: Series A (Statistics
  in Society)\/}~{\em 168\/}(1), 109--126.

\bibitem[\protect\citeauthoryear{Jeffreys}{Jeffreys}{1998}]{jeffreys1998theory}
Jeffreys, H. (1998).
\newblock {\em Theory of Probability}.
\newblock Oxford University Press.

\bibitem[\protect\citeauthoryear{Kang, Zhang, Cai, and Small}{Kang
  et~al.}{2016}]{kang2016instrumental}
Kang, H., A.~Zhang, T.~T. Cai, and D.~S. Small (2016).
\newblock Instrumental variables estimation with some invalid instruments and
  its application to {M}endelian randomization.
\newblock {\em Journal of the American Statistical Association\/}~{\em
  111\/}(513), 132--144.

\bibitem[\protect\citeauthoryear{Kass and Raftery}{Kass and
  Raftery}{1995}]{kass1995bayes}
Kass, R.~E. and A.~E. Raftery (1995).
\newblock Bayes factors.
\newblock {\em Journal of the American Statistical Association\/}~{\em
  90\/}(430), 773--795.

\bibitem[\protect\citeauthoryear{Kato et~al.}{Kato
  et~al.}{2013}]{kato2013quasi}
Kato, K. et~al. (2013).
\newblock Quasi-bayesian analysis of nonparametric instrumental variables
  models.
\newblock {\em The Annals of Statistics\/}~{\em 41\/}(5), 2359--2390.

\bibitem[\protect\citeauthoryear{Kozubowski and Podg{\'o}rski}{Kozubowski and
  Podg{\'o}rski}{2001}]{kozubowski2001asymmetric}
Kozubowski, T.~J. and K.~Podg{\'o}rski (2001).
\newblock Asymmetric laplace laws and modeling financial data.
\newblock {\em Mathematical and Computer Modelling\/}~{\em 34\/}(9--11),
  1003--1021.

\bibitem[\protect\citeauthoryear{Leamer}{Leamer}{1983}]{leamer1983let}
Leamer, E. (1983).
\newblock Let's take the con out of econometrics.
\newblock {\em American Economic Review\/}~{\em 73\/}(1), 31--43.

\bibitem[\protect\citeauthoryear{Li and Jiang}{Li and
  Jiang}{2016}]{li2016oracle}
Li, C. and W.~Jiang (2016).
\newblock On oracle property and asymptotic validity of bayesian generalized
  method of moments.
\newblock {\em Journal of Multivariate Analysis\/}~{\em 145}, 132--147.

\bibitem[\protect\citeauthoryear{Liao, Jiang, et~al.}{Liao
  et~al.}{2011}]{liao2011posterior}
Liao, Y., W.~Jiang, et~al. (2011).
\newblock Posterior consistency of nonparametric conditional moment restricted
  models.
\newblock {\em The Annals of Statistics\/}~{\em 39\/}(6), 3003--3031.

\bibitem[\protect\citeauthoryear{Lopes and Polson}{Lopes and
  Polson}{2014}]{lopes2014bayesian}
Lopes, H.~F. and N.~G. Polson (2014).
\newblock Bayesian instrumental variables: priors and likelihoods.
\newblock {\em Econometric Reviews\/}~{\em 33\/}(1-4), 100--121.

\bibitem[\protect\citeauthoryear{Madigan and Raftery}{Madigan and
  Raftery}{1994}]{madigan1994model}
Madigan, D. and A.~E. Raftery (1994).
\newblock Model selection and accounting for model uncertainty in graphical
  models using occam's window.
\newblock {\em Journal of the American Statistical Association\/}~{\em
  89\/}(428), 1535--1546.

\bibitem[\protect\citeauthoryear{Phillips}{Phillips}{1980}]{phillips1980exact}
Phillips, P. C.~B. (1980).
\newblock The exact distribution of instrumental variable estimators in an
  equation containing n+1 endogenous variables.
\newblock {\em Econometrica\/}~{\em 48\/}(4), 861--878.

\bibitem[\protect\citeauthoryear{Rubin}{Rubin}{1974}]{rubin1974estimating}
Rubin, D.~B. (1974).
\newblock Estimating causal effects of treatments in randomized and
  nonrandomized studies.
\newblock {\em Journal of Educational Psychology\/}~{\em 66\/}(5), 688--701.

\bibitem[\protect\citeauthoryear{Rubin}{Rubin}{1978}]{rubin1978bayesian}
Rubin, D.~B. (1978).
\newblock Bayesian inference for causal effects: The role of randomization.
\newblock {\em The Annals of statistics\/}~{\em 6}, 34--58.

\bibitem[\protect\citeauthoryear{Small}{Small}{2007}]{small2007sensitivity}
Small, D.~S. (2007).
\newblock Sensitivity analysis for instrumental variables regression with
  overidentifying restrictions.
\newblock {\em Journal of the American Statistical Association\/}~{\em
  102\/}(479), 1049--1058.

\bibitem[\protect\citeauthoryear{Smith and Ebrahim}{Smith and
  Ebrahim}{2003}]{davey2003mendelian}
Smith, G.~D. and S.~Ebrahim (2003).
\newblock `mendelian randomization': can genetic epidemiology contribute to
  understanding environmental determinants of disease?
\newblock {\em International {J}ournal of {E}pidemiology\/}~{\em 32\/}(1),
  1--22.

\bibitem[\protect\citeauthoryear{Smith and Ebrahim}{Smith and
  Ebrahim}{2004}]{smith2004mendelian}
Smith, G.~D. and S.~Ebrahim (2004).
\newblock Mendelian randomization: prospects, potentials, and limitations.
\newblock {\em International Journal of Epidemiology\/}~{\em 33\/}(1), 30--42.

\bibitem[\protect\citeauthoryear{Windmeijer, Farbmacher, Davies, and
  Davey~Smith}{Windmeijer et~al.}{2018}]{windmeijer2018use}
Windmeijer, F., H.~Farbmacher, N.~Davies, and G.~Davey~Smith (2018).
\newblock On the use of the lasso for instrumental variables estimation with
  some invalid instruments.
\newblock {\em Journal of the American Statistical Association\/}, 1--12.

\bibitem[\protect\citeauthoryear{Yin et~al.}{Yin
  et~al.}{2009}]{yin2009bayesian}
Yin, G. et~al. (2009).
\newblock Bayesian generalized method of moments.
\newblock {\em Bayesian Analysis\/}~{\em 4\/}(2), 191--207.

\end{thebibliography}
\end{document}